\documentclass[]{aastex631}

\begin{document}

\title{Imaging a large coronal loop using type U solar radio burst interferometry}

\author[0000-0001-8641-9627]{Jinge Zhang}
\affiliation{Mullard Space Science Laboratory, University College London, RH5 6NT, United Kingdom}
\email{jinge.zhang.18@ucl.ac.uk}

\author[0000-0002-6287-3494]{Hamish A.S. Reid}
\affiliation{Mullard Space Science Laboratory, University College London, RH5 6NT, United Kingdom}

\author{Eoin Carley}
\affiliation{Astronomy \& Astrophysics Section, Dublin Institute for Advanced Studies, Dublin D02 XF86, Ireland}

\author[0000-0002-8428-1369]{Laurent Lamy}
\affiliation{LESIA, Observatoire de Paris, Universit\'e PSL, CNRS, Sorbonne Universit\'e, Universit\'e de Paris, Meudon, France}
\affiliation{Observatoire de Radioastronomie de Nan\c cay, Observatoire de Paris, Universit\'e PSL, CNRS, Univ. Orl\'eans, Nan\c cay, France}
\affiliation{Aix Marseille Univ, CNRS, CNES, LAM, Marseille, France}

\author[0000-0002-6760-797X]{Pietro Zucca}
\affiliation{ASTRON, The Netherlands Institute for Radio Astronomy, Oude Hoogeveensedijk 4, 7991 PD Dwingeloo, The Netherlands}

\author[0000-0001-6855-5799]{Peijin Zhang}
\affiliation{Center for Solar-Terrestrial Research, New Jersey Institute of Technology, Newark, NJ, USA}
\affiliation{Cooperative Programs for the Advancement of Earth System Science, University Corporation for Atmospheric Research, Boulder, CO, USA}

\author[0000-0001-7915-5571]{Baptiste Cecconi}
\affiliation{LESIA, Observatoire de Paris, Universit\'e PSL, CNRS, Sorbonne Universit\'e, Universit\'e de Paris, Meudon, France}
\affiliation{Observatoire de Radioastronomie de Nan\c cay, Observatoire de Paris, Universit\'e PSL, CNRS, Univ. Orl\'eans, Nan\c cay, France}

\begin{abstract}

Solar radio U-bursts are generated by electron beams traveling along closed magnetic loops in the solar corona. Low-frequency ($<$ 100 MHz) U-bursts serve as powerful diagnostic tools for studying large-sized coronal loops that extend into the middle corona. However, the positive frequency drift component (descending leg) of U-bursts has received less attention in previous studies, as the descending radio flux is weak.  In this study, we utilized LOFAR interferometric solar imaging data from a U-burst that has a significant descending leg component, observed between 10 to 90 MHz on June 5th, 2020.  By analyzing the radio source centroid positions, we determined the beam velocities and physical parameters of a large coronal magnetic loop that reached just about 1.3 $\rm{R_{\odot}}$ in altitude.  At this altitude, we found the plasma temperature to be around 1.1 MK, the plasma pressure around 0.20 $\rm{mdyn,cm^{-2}}$, and the minimum magnetic field strength around 0.07 G. The similarity in physical properties determined from the image suggests a symmetric loop. The average electron beam velocity on the ascending leg was found to be 0.21 c, while it was 0.14 c on the descending leg. This apparent deceleration is attributed to a decrease in the range of electron energies that resonate with Langmuir waves, likely due to the positive background plasma density gradient along the downward loop leg. 

\end{abstract}



\section{Introduction} \label{sec:intro}

The solar type U burst, commonly called U-burst, was firstly reported by \citet{Maxwell1958} as these bursts were observed as inverted "U" shapes on dynamic spectra. U-bursts are generally believed to be generated by accelerated electron beams propagating along closed magnetic loops in the solar corona \citep[e.g.,][]{Klein1993, Karlicky1996}. Explained by the theory of plasma emission mechanism based on the work of \citet{Ginzburg1959}, the observational frequencies ($f_{\text{obs}}$) of a U-burst are equal to or double the background electron plasma frequencies ($f_{e}$) along the coronal loop where the electron beam travels, if they are fundamental or second-harmonic emissions, respectively. Although the solar corona is an inhomogeneous medium, it is typical for the corona plasma density to decrease as the altitude increases. A U-burst can serve as a powerful diagnostic tool to probe the physical parameters of closed coronal magnetic loops and to study the kinetic properties of the accelerated electron beams traveling in these loops, located in the higher region of the corona, such as the middle corona.

The middle corona, defined by heliocentric distances from 1.5 to 6 solar radii ($\rm{R_{\odot}}$), has a tenuous plasma environment where the density is too low for X-ray and EUV emission diagnostics \citep[see][]{West2023}. The low-frequency U-bursts, observed by ground-based radio interferometers below 100 MHz, are generated by electron beams traveling along large-sized coronal loops that extend into the middle corona, higher than 1.5 $\rm{R_{\odot}}$ heliocentric height. These large sized closed magnetic loop structures have been poorly studied due to the historical lack of low-frequency radio imaging, which was present only at a few, discrete frequencies. For example, \citet{Stewart1977} used 80 and 130 MHz from the Culgoora Radioheliograph.  Analysing the negative frequency drift component (ascending leg) and positive frequency drift component (descending leg) of a low-frequency U-burst with high frequency resolution (~1 MHz) radio interferometric imaging, such as the capability provided by the LOw-Frequency-ARray (LOFAR) \citep[see][]{vanHaarlem2013}, can determine the velocity of electron beams traveling along both legs of the loop. It can also be used to determine the physical parameters of such large-sized loops, which have historically been ill-defined.

Previous studies have estimated the velocity of electron beams from the ascending leg of U-bursts, where electron beam propagating outwards from the Sun. Using methods similar to those employed in many previous studies on type III bursts \citep[e.g.,][]{Poquerusse1994,Mann1999,Reid2017}, one can convert observed frequency to plasma frequency using $f_{\text{obs}} 
 \approx f_{e}$ for fundamental emission or $f_{\text{obs}} 
 \approx 2f_{e}$ for second-harmonic emission. 
The ambient electron density $n_e$ can be derived. Using a numerical coronal density model, the radial distance of each radio source along the coronal magnetic flux tube, $l$, can be determined. Therefore, the evolution of frequency in time from a type III burst or the ascending leg of a U-burst can thus provide the electron beam velocity \citep[This method has been summarized by][]{Reid2014}. \citet{Labrum1970} estimated an average beam velocity of 0.25c by analyzing 29 U-bursts below 160 MHz, assuming fundamental emission and Newkirk's coronal density model. Recently, \citet{Zhang2023} determined the electron beam velocity from 27 type J bursts between 20 to 80 MHz observed by LOFAR during the same solar radio noise storm, averaging at 0.22c by assuming second-harmonic emission and multiplying Saito's density model\citep[the background coronal density model by][]{Saito1977} by a factor of 4.5. However, this method cannot be applied to the turning over part and the descending leg of U-bursts because the standard coronal density models cannot be used when the magnetic field lines bend horizontally.

Using solar radio imaging measurements to directly determine the position of radio sources for predicting electron beam velocities and coronal loop parameters has become popular in recent years \citep[see][as a recent review]{Reid2020}, thanks to the construction and operation of high-performance new generation radio telescopes, such as LOFAR. By using the LOFAR tied-array beams solar images, \citet{Reid2017} estimated the exciter velocities of three U-bursts while they traveled along ascending legs between 40-70 MHz, ranging from 0.16c to 0.20c. However, determining the electron beam velocities from the descending leg was difficult due to the weak and faint emissions across narrow frequency ranges. Similarly, \citet{Dabrowski2023} analyzed the velocities of type-III and U bursts in their study using LOFAR tied-array and interferometric imaging. However, they were only able to determine the exciter velocity of the selected U-burst from the dynamic spectrum, as the extremely diffuse flux in the descending leg made imaging the descending leg challenging, especially in the lower frequency range below 100 MHz. Therefore, imaging low frequency U-bursts with clear and strong descending leg emissions would provide better opportunities for studying the propagation of electron beams along the downward leg of the coronal loops.

Other studies have used U-bursts to determine the physical parameters of coronal loops \citep[e.g.,][]{Aschwanden1992,Fernandes2012}. Using GHz radio observations from the Very Large Array (VLA), \citet{Aschwanden1992} derived the density, temperature, pressure, and minimum magnetic field strength of three loops around 20 Mm in altitude. At MHz frequencies, using the NRH data, \citet{Mancuso2023} determined coronal loop parameters by constructing the 3D morphology of the loop from five radio source positions on images at three different frequencies (360.8, 327.0, 298.7 MHz). This loop reached about 146 Mm (0.21 solar radii) in altitude. Due to the limited number of frequency bands available for imaging, deriving loop parameters for the ascending and descending leg individually becomes challenging. However, coronal loops that generate U-bursts are commonly assumed to possess symmetrical geometric and physical properties. It remains unclear whether the ascending and descending legs exhibit similar or different physical characteristics, due to the lack of adequate simultaneous measurements of both legs for comparison.

In this work, we study an extremely bright U-burst observed by LOFAR between 10 to 90 MHz on the 5th June 2020. The U-burst has clear descending leg structure from 20 to 40 MHz, which provides opportunities for imaging the descending leg sources positions by using LOFAR interferometric imaging with high frequency resolutions. We estimated electron beam velocities travelled along upward and downward flows based on ascending and descending legs images. We derived coronal loops parameters by inferring plasma density models along the closed loop from radio images. The present paper is organized as follows. In Section 2, we describe the details of the U-burst, including the LOFAR dynamic spectrum and observations from Nan\c cay Decameter Array (NDA) and e-Callisto, as they provided additional information about the event from LOFAR. In Section 3, we present the LOFAR radio interferometric image data of the U-burst. We plotted the coronal loop structure over the solar disc and report the radio source characteristics of the U-burst, including source sizes, position, and time. In Section 4, we determine the electron beam velocities from both the ascending and descending legs on the image to make comparisons, as well as derive the coronal loop parameters. The discussions of the results are presented in Section 5 to address the questions mentioned in this section, including discussions of the scattering and projection effects. The final section concludes the key findings of this paper, and we also suggest future studies of large-size coronal loops and their connections to the low corona and interplanetary space by utilizing U-burst events.

\section{Observations of the U-burst} \label{sec:OBS}

\begin{figure}[ht!]
\plotone{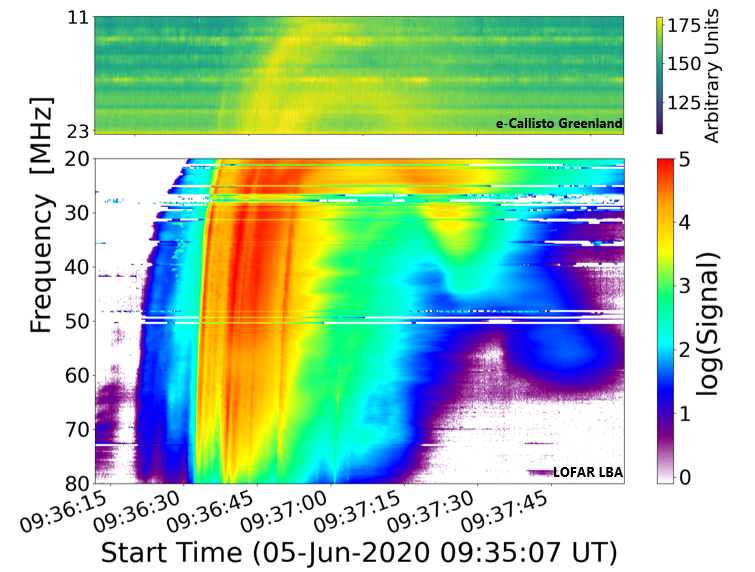}
\caption{
Dynamic spectrum of the U-burst event on June 5th 2020. The bottom panel displays the LOFAR dynamic spectrum from 20 to 80 MHz, while the upper panel shows the e-Callisto dynamic spectrum between 11 to 24 MHz. The turning frequency is approximately 18 MHz at 09:37:00 UT. 
\label{fig:DynSpec}}
\end{figure}

The U-burst was observed by several ground-based radio telescopes at 09:37:00 UT on June 5th, 2020, between 10 to 90 MHz. \citet{Stanislavsky2021, Stanislavsky2022} reported that the Giant Ukrainian Radio Telescope (GURT), the NDA, the Radio Solar Telescope Network (RSTN), and the e–Callisto Greenland sites observed the event simultaneously. On the LOFAR dynamic spectrum shown in Figure \ref{fig:DynSpec}, it is clear that the event contains a group of type-III bursts preceding and following the ascending leg flux of the U-burst, which represents the negative frequency drift rate leg. This makes it challenging to determine which branch corresponds to the ascending leg flux of the U-burst.

The bright radio flux in the event was observed starting around 09:36:32 UT, consisting of a group of type-III-like bursts (which may also contain J-bursts) that emerged together during a short period of time until 09:36:50 UT. This group of bursts contains the ascending leg of the U-burst. However, the ascending leg flux overlaps with a bright type III burst at 09:36:45 UT on the LOFAR dynamic spectrum. Together, this branch of the flux exhibits the largest temporal width on the dynamic spectrum. As evidence, the e-Callisto dynamic spectrum in Figure \ref{fig:DynSpec} shows this flux branch separating around 23 MHz into two.  The first flux branch drifts towards lower frequencies (around 11 MHz), overlaps with the preceding bright type III burst that starts around 09:36:40 UT and exhibits a relatively constant frequency drift rate.  We consider this to be a type-III burst. The second flux branch has a clear characteristic of decreased drift rate, reached the turning over point at around 18 MHz on the e-Callisto dynamic spectrum. Therefore, we have identified the second separated branch as the ascending leg flux of the U-burst. The descending leg of the U-burst is defined as the emissions occurring after the turning point with a positive frequency drift rate. On the LOFAR dynamic spectrum, it can be clearly identified as the flux drifting from 20 to 40 MHz, between 09:37:14 and 09:37:26 UT.

\begin{figure}[ht!]
\plotone{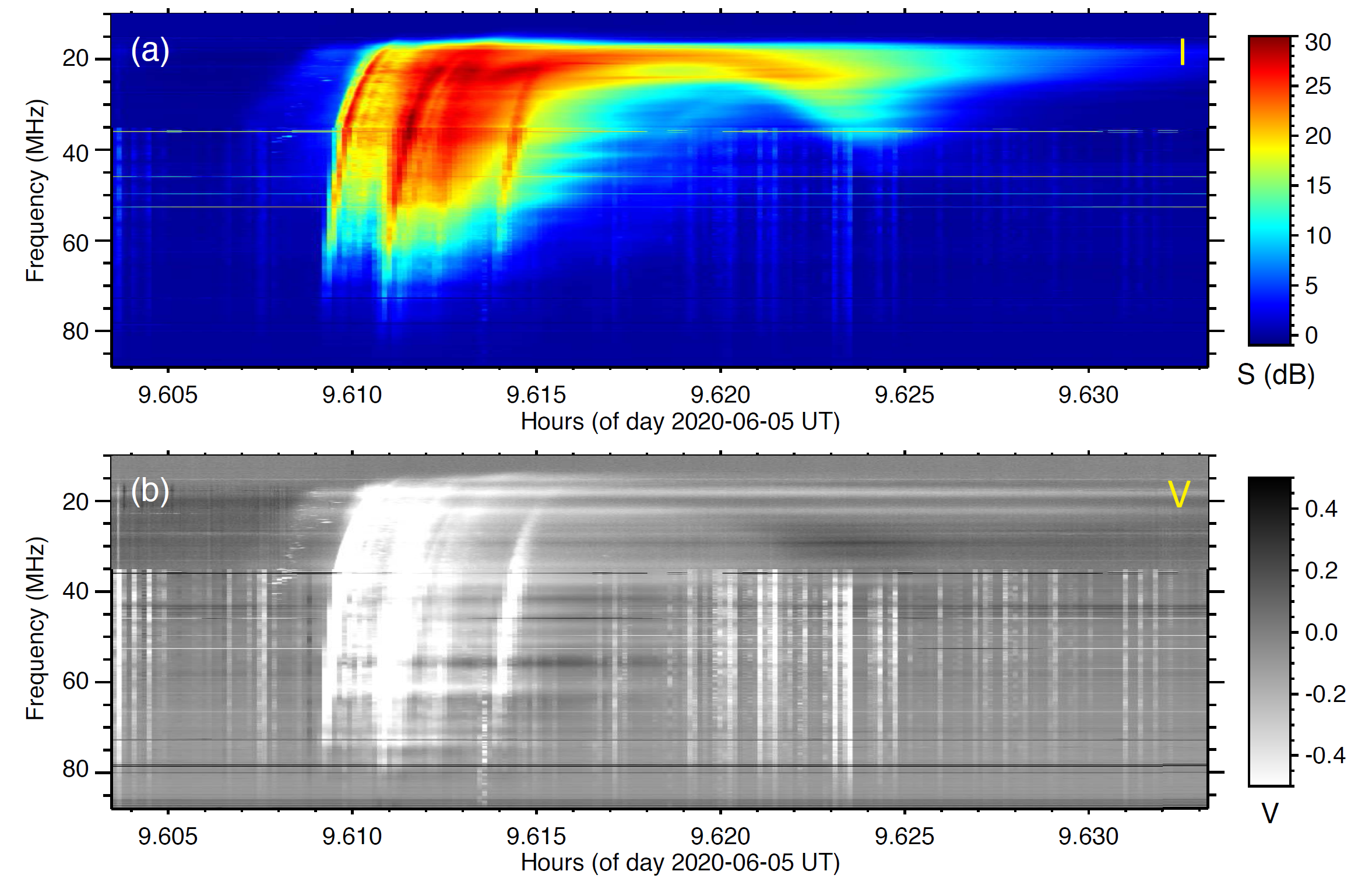}
\caption{
NDA composite dynamic spectrum of Stokes I (top) and Stokes V (bottom) using the NewRoutine receiver above 35~MHz and the Mefisto one (RFI-filtered data) below 35 MHz. The type III bursts were highly right-handed circularly polarized ($|V| > 0.4$). In comparison, the U-burst has a lower degree of circular polarization ($|V| < 0.2$).  The ascending leg of the U-burst is right-handed circularly polarized whilst the decending leg is left-hand circularly polarised.
\label{fig:NDA}}
\end{figure}

Figure \ref{fig:NDA} displays a NDA composite dynamic spectrum of the Stokes V parameter \citep{Boischot1980}. It merges the data obtained from the NewRoutine and Mefisto (which filters the data on-the-fly) receivers above and below 35~MHz, respectively \citep{Lamy2017,Lecacheux13}. We observe that the degree of circular polarization was different for the U-burst and the surrounding visible type III bursts. Most of the type III bursts were highly right-handed circularly polarized. The U-burst ascending leg displayed a relatively lower degree of right-handed circular polarization ($|V|/I < 0.2$), which entirely switched to left-handed circular polarization for the U-burst descending leg. The U-burst lower degree of circular polarization allows us to confidently conclude that the emission mechanism is second-harmonic emission, which typically produces low circular polarized radio flux \citep[see][]{Dulk1980}. In addition, the switch in the sign of V between the ascending and descending legs (from RH to LH) is consistent with the fact that the electron beam is traveling in opposite directions relative to the observer.

We also report the observation of another descending leg component by LOFAR in this event, which is visible on the log-scale LOFAR dynamic spectrum, as shown in Figure \ref{fig:DynSpec}. The radio flux was recorded between 40 and 60 MHz, from 09:37:40 to 09:38:00 UT. The reason for identifying it as another U-burst rather than a continuation of the U-burst in this study is because it exhibits a lower frequency drift rate of 1.6 MHz/s,  lower than the U-burst descending leg drift rate at 2.3 MHz/s, and two order of magnitude lower in intensity. It is challenging to identify the turning point and ascending leg of this U-burst as they were overshadowed by other bright radio bursts, similar to the main U-burst ascending leg that we are studying.

\section{LOFAR Interferometric Imaging}\label{sec:Imaging}

The LOFAR interferometric solar imaging dataset consists of imaging data from 60 frequency sub-bands, ranging from 19.13 to 80.07 MHz, with a step of approximately 1~MHz, varying slightly for each sub-band.

\subsection{The U-burst Image}\label{sec:U-burst image}

\begin{figure}[ht!]
\plotone{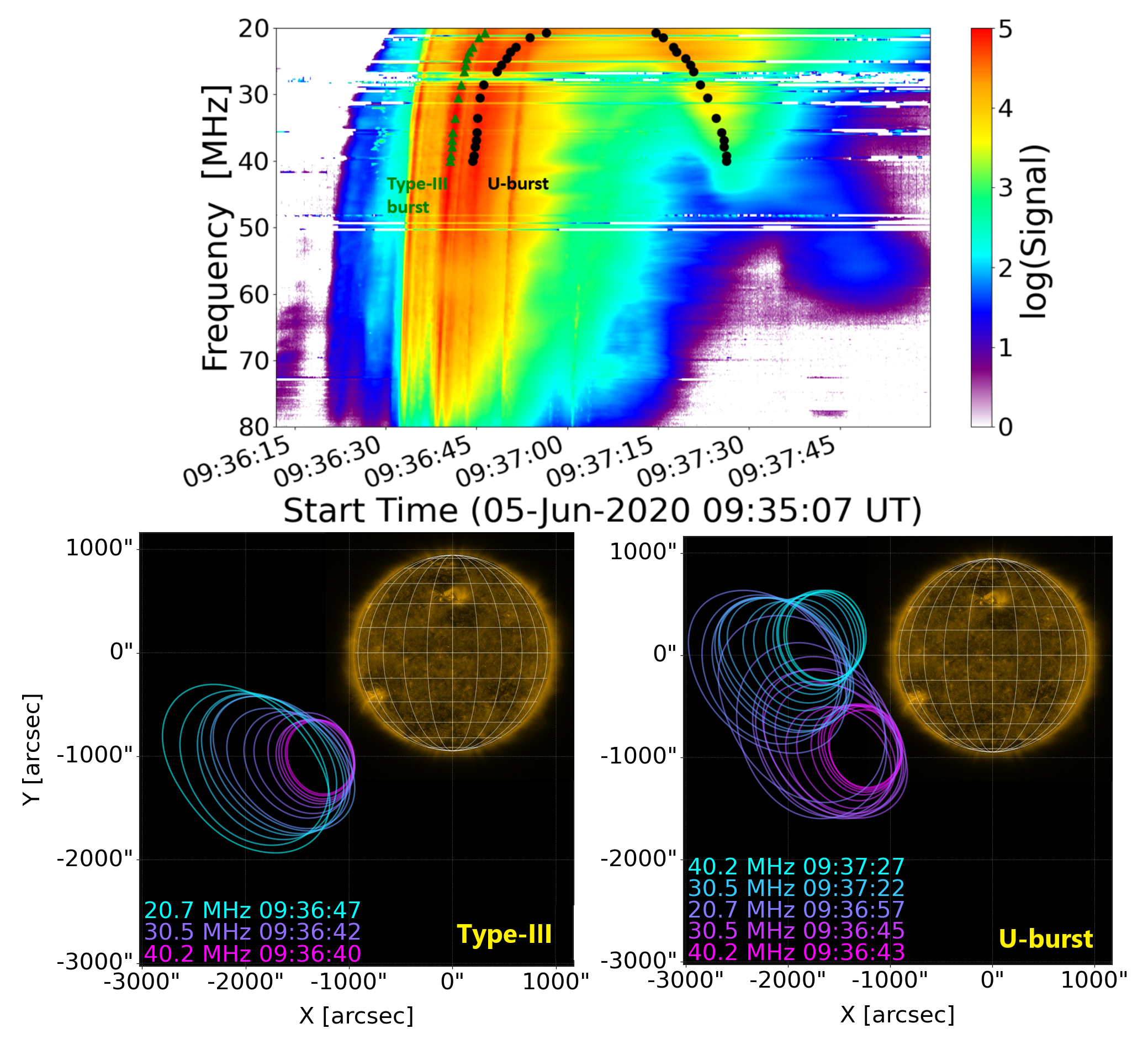}
\caption{Top: dynamic spectrum of the event.  The times of peak flux for each image contour are highlighted by green triangles for the type III burst and black dots for the U-burst.  Bottom: LOFAR interferometric solar radio image contours taken at 50\% intensity for a type III burst (left) and the U-burst (right), between 20-40 MHz.  Image contours are overlaid on the AIA 171 Å image.  
\label{fig:TheImg}}
\end{figure}

Since the descending leg of the U-burst stopped emitting at 40 MHz, we primarily utilized the imaging data between 20.70 MHz and 40.03 MHz to generate the radio source image of the U-burst. The temporal resolution of the imaging data for each frequency sub-band is approximately 1.7 seconds, after being processed using the LOFAR solar image pipeline. In total, we selected 15 noiseless frequency sub-bands to create the interferometric image of the U-burst. For each selected sub-band, two time intervals were chosen to represent the ascending and descending legs respectively, as black dots labeled on the dynamic spectrum, shown in Figure \ref{fig:TheImg}. The time profile for the ascending leg was determined based on the peak flux time during the ascending flux period. The time profile for the descending leg can be directly identified on the descending flux branch as the peak flux occurring between 09:37:13 and 09:37:27 UT. Therefore, we have 30 radio images to visualize the propagation of the electron beam along the closed magnetic loops that generated the U-burst. These radio images, overplotted on the AIA 171 Å image of the Sun, depict the propagation of the electron beam along the large coronal magnetic loop, as shown in Figure \ref{fig:TheImg}. In this image, the contours are plotted at the full width at half maximum (FWHM) representing the sizes of the radio sources along the loop.

\begin{figure}[ht!]
\plotone{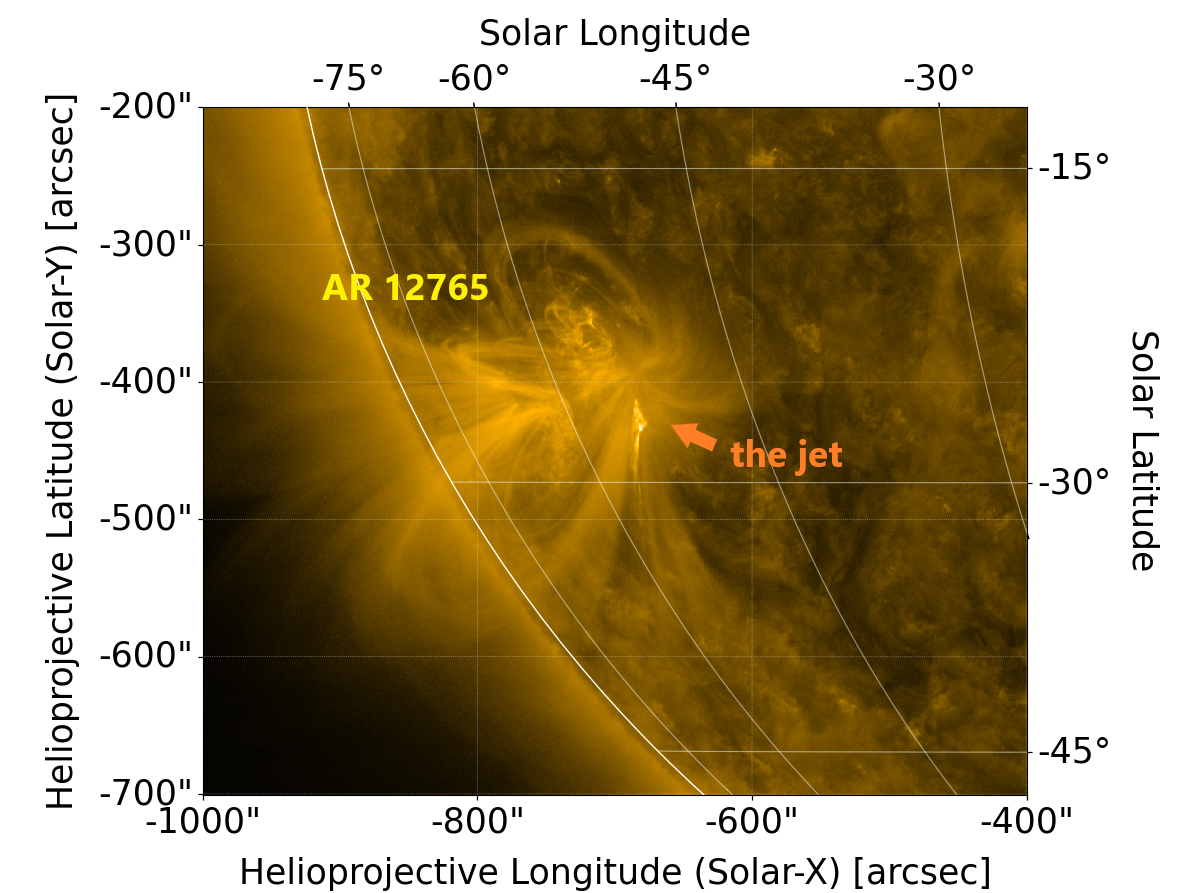}
\caption{
AIA 171 Å image of an observed coronal jet that was co-temporal with the U-burst, originating from Active Region AR12765 at 2020-06-05 09:37:45 UT.
\label{fig:JET}}
\end{figure}

The magenta contours on the U-burst image illustrate the ascending leg of the U-burst, collectively demonstrating the outward travel of the electron beam from the Sun. The highest-frequency channel selected for imaging the ascending leg of the U-burst is 40.2 MHz, located at a high altitude off the solar disc in the image. This positioning makes it challenging to detect the origin of the electron beam acceleration, as no GOES X-ray activities have been observed. However, a coronal jet was observed simultaneously from AR12765 during the U-burst event, as shown in Figure \ref{fig:JET}. We consider that the accelerated electron beams that produced the radio emissions of this U-burst event were accompanied by this jet plasma and they were accelerated by the same magnetic reconnection process. This acceleration process produced a minimal amount of emissions compared to typical flares \citep[see][]{Raouafi2016}. The observation of this jet provides evidence supporting that the footpoint of the upward leg of the large coronal loop was located at AR12765. 

\cite{Chen2023} investigated the source positions and directivity of the interplanetary (IP) type III bursts, which were simultaneously observed below 10 MHz by the Parker Solar Probe, Solar Orbiter, and STEREO. We established that this interplanetary type III burst was the same event as the U-burst in our study, as it was connected to the type III burst flux we imaged in Figure \ref{fig:TheImg}. This connection was evidenced by the clear extension of flux to the lower frequency range on the e-Callisto dynamic spectrum, as shown in Figure \ref{fig:DynSpec}. \cite{Chen2023} employed both intensity fitting and timing methods (indicated by triangle symbols) to determine the positions of the interplanetary type III burst sources. Furthermore, they confirmed that the electron beam originated from AR12765 and that its trajectory was straightforward along the -60° longitudinal plane.

Other type III bursts observed during the ascending period of the U-burst event also originated from this active region. As an example, we imaged the brightest preceding type III burst, as shown in Figure \ref{fig:TheImg}; the high-frequency radio sources of the type III burst are spatially close to the ascending leg sources of the U-burst. In this study, we assumed that the ascending leg sources were located on the same longitudinal plane (-60°) as the active region AR12765 at 09:37:00 UT.

The sources in the ascending leg, ranging from 30.5 to 20.7 MHz on the image, depict the electron beam's travel along the upward leg of the large coronal loop drifted towards the north and eventually reached the loop apex above the solar equatorial region. This observation aligns well with the general understanding of the U-burst emission mechanism, which places the emission region of the turning over part of the burst at the loop apex. The radio sources at the lowest frequency of both the ascending and descending legs, observed at 20.7 MHz, exhibit a noticeable shift towards the north. The region in which the loop apex can be considered located is suggested to be within this area based on the image.

\begin{figure}[ht!]
\plotone{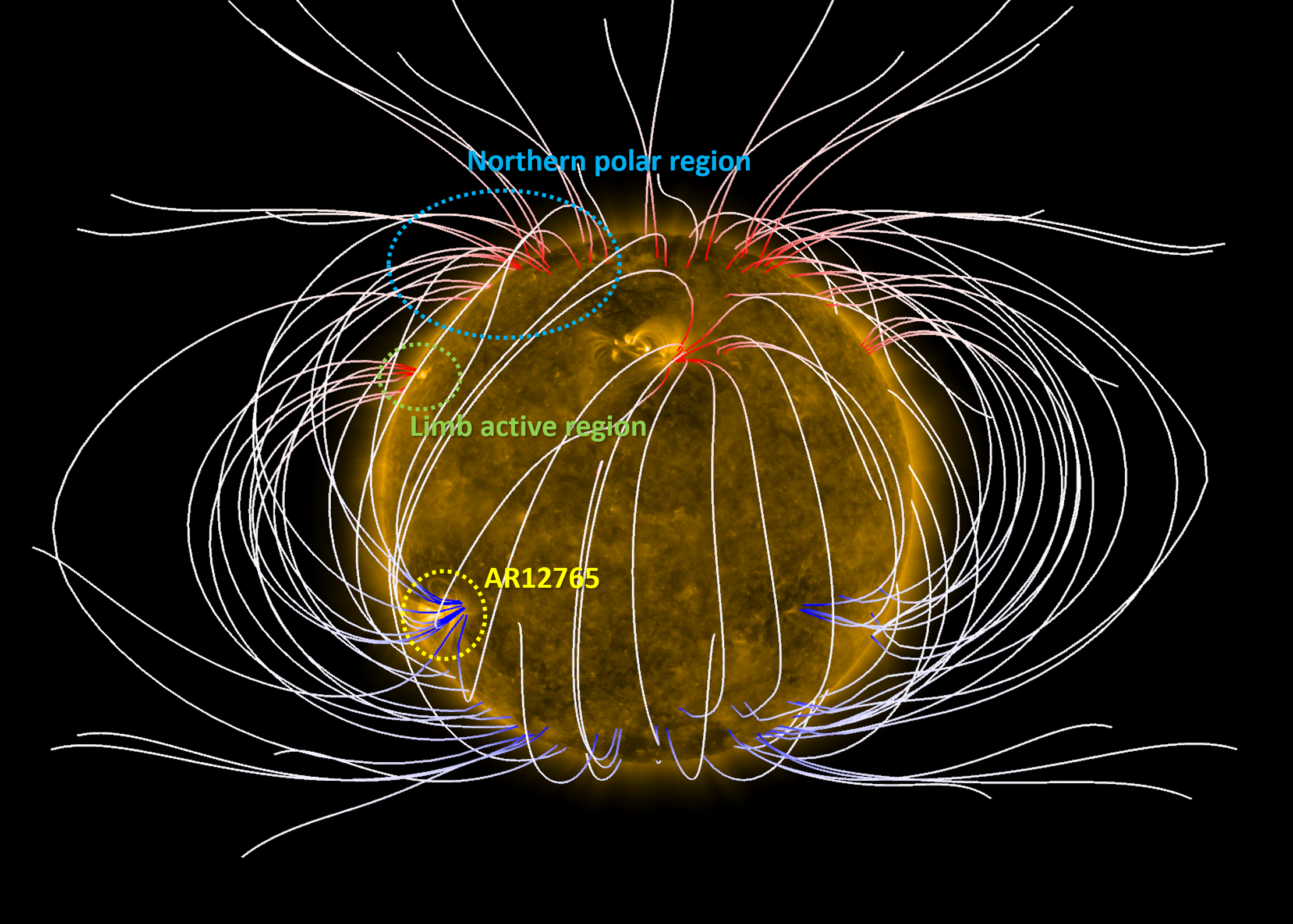}
\caption{
Potential Field Source Surface (PFSS) global modelling of the solar magnetic field on 5th June 2020 09:06:37 UT, with the source surface set at 2.5 $\rm{R_{\odot}}$.
\label{fig:PFSS}}
\end{figure}

The radio sources of the descending leg, indicated by cyan contours on the U-burst image in Figure \ref{fig:TheImg}, show that the electron beam is propagating downward from the apex of the loop above the solar equatorial region towards the high-latitude regions in the northern solar hemisphere. This observation suggests that the location of the footpoint for the downward leg of the magnetic loop is in the northern high-latitude regions. Regarding the Potential Field Source Surface (PFSS) global modeling, with the source surface set at 2.5 $\rm{R_{\odot}}$ as illustrated in Figure \ref{fig:PFSS}, there are large magnetic field loops connecting AR12765 to various northern high-latitude regions. Likely locations include the southern polar region, the unidentified active region located near the limb, and some field lines connected to the backside of the solar limb. However, none of these loops have a 2D projection that reaches the heliocentric height depicted in the radio image, although the source surface height at 2.5 $\rm{R_{\odot}}$ is considered sufficient to display the corresponding large coronal loop from extrapolation. The uncertainty of the coronal loop downward leg footpoint location leads to the challenge of the projection effect, as the true position of radio sources is hard to measure due to the distances along the line-of-sight axis being undetectable. In this study, we assumed that the descending leg sources were located on the same plane as the solar limb for studying the coronal loop physical parameters and the beam velocity. We also discussed the projection effect's influence on the results in Section \ref{sec:Projection}.

\section{Electron Beam Characteristics and the Coronal Loop Physical Parameters}\label{sec:parameters}

\subsection{Coronal Loop Density Model}\label{sec:DEN&Vel}

The background electron density model of the coronal loop can be directly inferred from the U-burst images. The positions of the radio sources were defined as the centroids of each contour on the image. As explained in the previous section, to approximate projection effects, we assumed that the centroids of the ascending leg radio sources were located on the same longitudinal plane (-60°) as the active region AR12765 at 09:37:00 UT, and the descending leg sources were located on the same plane as the solar limb (-90°). The assumption of both longitudinal planes allows us to define the observational angle of each radio source in addition to the 2D projection radio images.  The ascending leg source height spanning from 0.8 to 1.3 $\rm{R_{\odot}}$, and the descending leg source height from 0.7 to 1.3 $\rm{R_{\odot}}$. Both legs are within the same range of the middle corona, with the average height of the ascending leg being only 3\% higher than that of the descending leg.

The electron density, $n_e$ can be determined from the radio frequency profile of the contours $f_r$, assuming the emission mechanism is second-harmonic, by:

\begin{equation}
\label{eqn:F2D}
n_\mathrm{e} = \frac{m_e (\pi f_r)^2}{4\pi e^2},
\end{equation}

where $m_e$ denotes the electron mass, $e$ represents the electron charge.

The background electron density distribution of the coronal loop can be described by fitting an exponential model: 
\begin{equation}  \label{eqn:exponential}
n_\mathrm{e}(h) = n_0 \exp{\left(\frac{-h}{\lambda}\right)},
\end{equation}  
Where $n_e$ is the electron plasma density at solar altitude $h$. $n_0$ denotes the reference plasma density, a constant determined from the exponential fit, and $\lambda$ is the hydrostatic plasma density scale height.

\begin{figure}[ht!]
\plotone{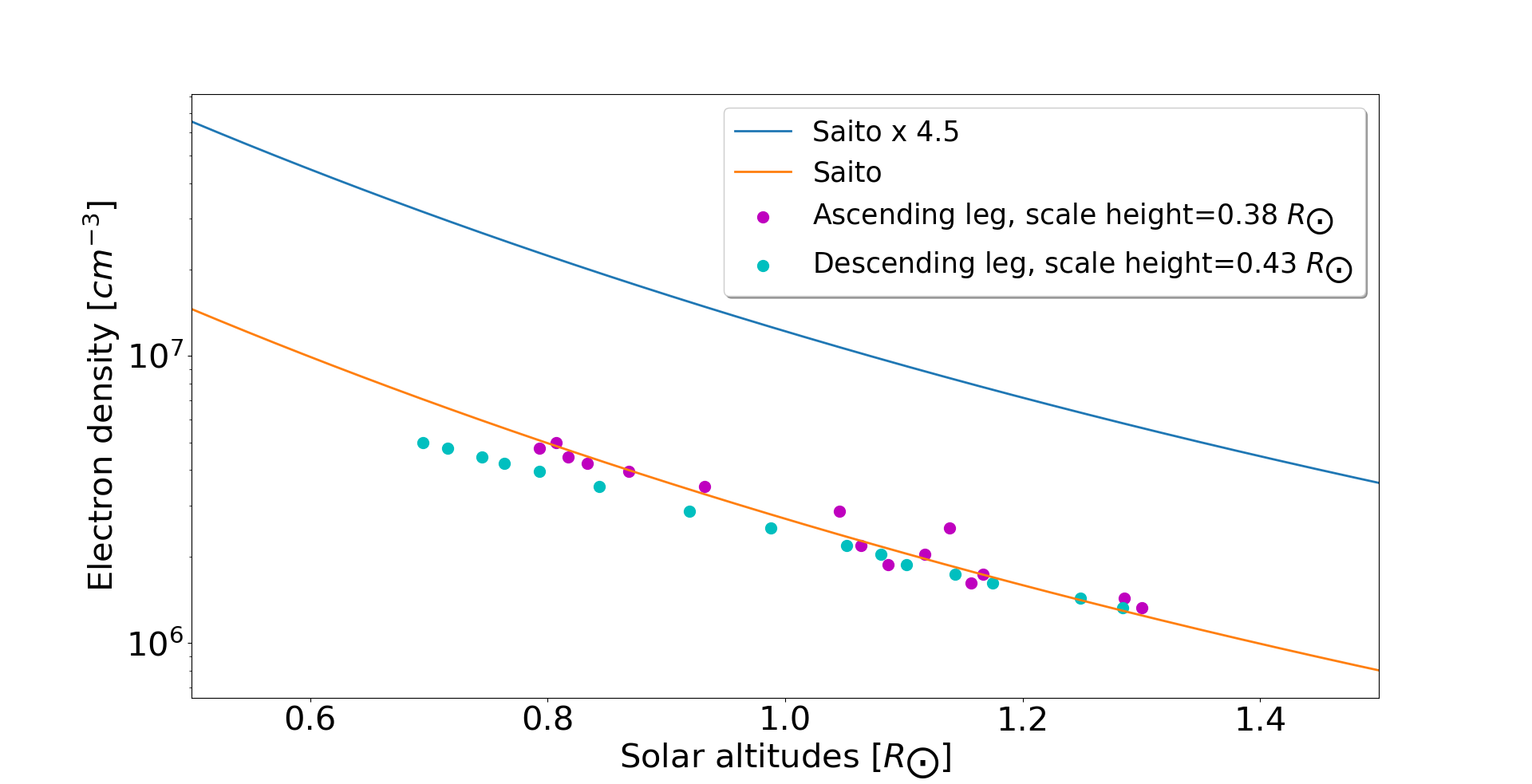}
\caption{Electron density as a function of solar altitude for the ascending leg (purple dots) and descending leg (cyan dots) of the coronal loop. Overplotted is the Saito density model (orange) and the Saito density model times 4.5, which was inferred by \citep{Reid2017}.
\label{fig:NRplot}}
\end{figure}

We found that the magnitude of the density models from both legs is similar to Saito's model, as shown in Figure \ref{fig:NRplot}. The density model of the descending leg exhibits a slightly lower gradient, which means a larger density scale height compared to the ascending leg. By fitting an exponential density model, we determined the density scale height, denoted as $\lambda$ in Equation \ref{eqn:exponential}, for both legs. The ascending leg density scale height, $\lambda_{ascending}= 0.38~\rm{R_{\odot}}$, and the descending leg scale height, $\lambda_{descending}= 0.43~\rm{R_{\odot}}$. There is only 12\% difference between the scale heights, likely within the errors given by projection and scattering effect.  We consider this as evidence of a symmetric loop in plasma conditions.  Both scale heights are close to the statistical result of 0.36 $\rm{R_{\odot}}$ from the analysis of 24 type J bursts observed in the same frequency range, as reported by \citet{Zhang2023}.

\subsection{Electron Beam Velocity}\label{sec:Velocity}

\begin{figure}[ht!]
\plotone{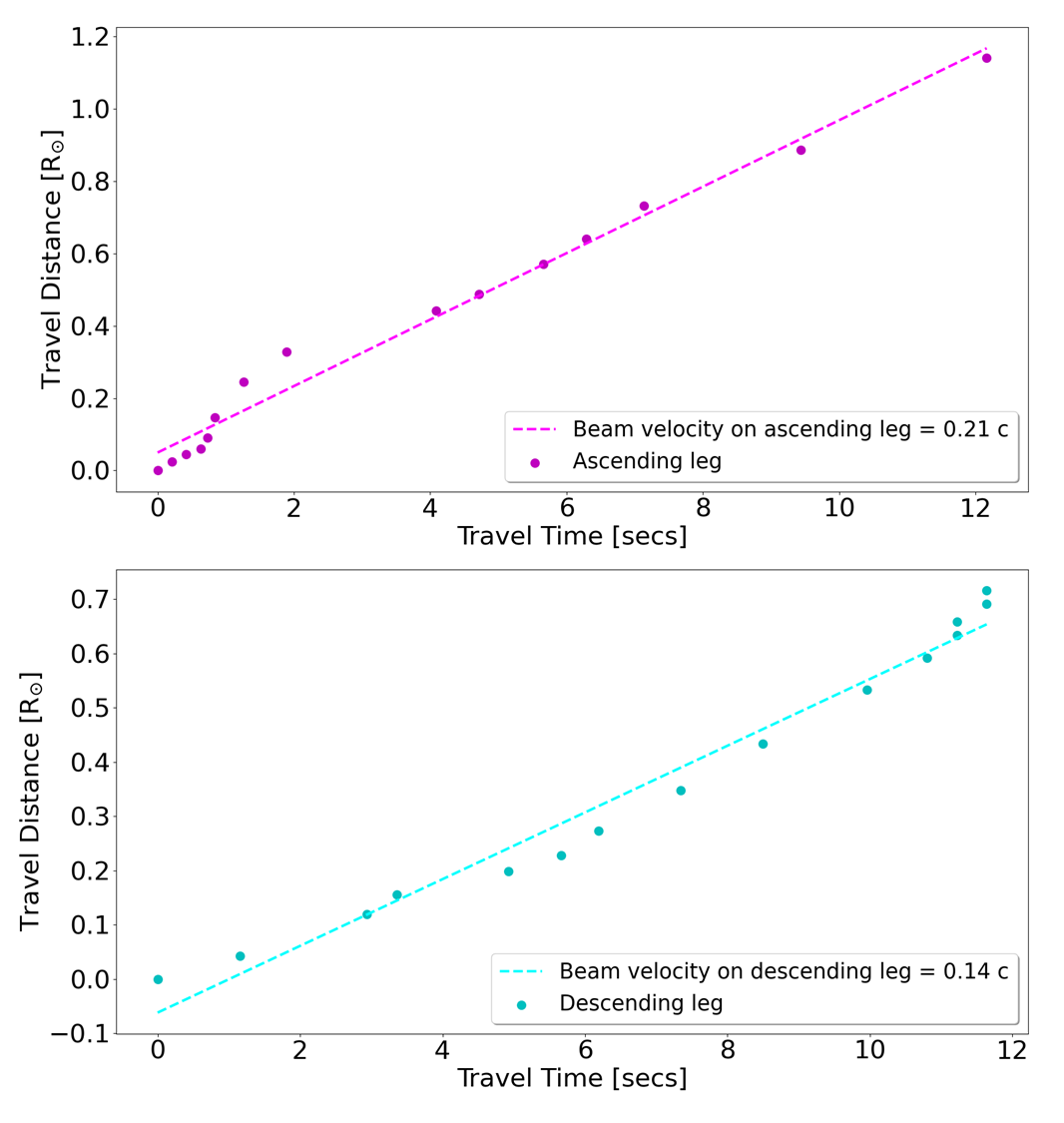}
\caption{
Distances between each centroid of the radio image contours vs travel time, determined for the ascending leg (top) and descending leg (bottom).  The 40.2 MHz contour centroid is the start point of the ascending leg, and the 20.7 MHz contour centroid is the start point of the descending leg. The travel distance is defined as the distances between each centroid to the previous one.  The electron beam velocity along the ascending leg (0.21c) was greater than that along the descending leg (0.14c).
\label{fig:RT_Plots}}
\end{figure}

We determined the electron beam velocity directly from the image by analyzing the position of the radio source, which reflected the actual path of the propagating electron beam along with a specific travel time profile. To calculate the electron beam velocity, we measured the travel distance as the sum of the distances between each centroid position. The travel time was defined as the peak flux point time profile selected from the LOFAR dynamic spectrum. The relationship between travel distance and travel time is illustrated in Figure \ref{fig:RT_Plots}.

The electron beam traveled both legs with similar travel times. Specifically, the ascending leg exhibited a travel time of 12.16 seconds, while the descending leg had a travel time of 11.64 seconds. The total travel distance for the ascending leg is 1.14 $\rm{R_{\odot}}$, and for the descending leg, it is 0.72 $\rm{R_{\odot}}$. By examining the plot in Figure \ref{fig:RT_Plots}, we observed that the electron beam velocity along the ascending leg (0.21c) was greater than that along the descending leg (0.14c).

\begin{figure}[ht!]
\plotone{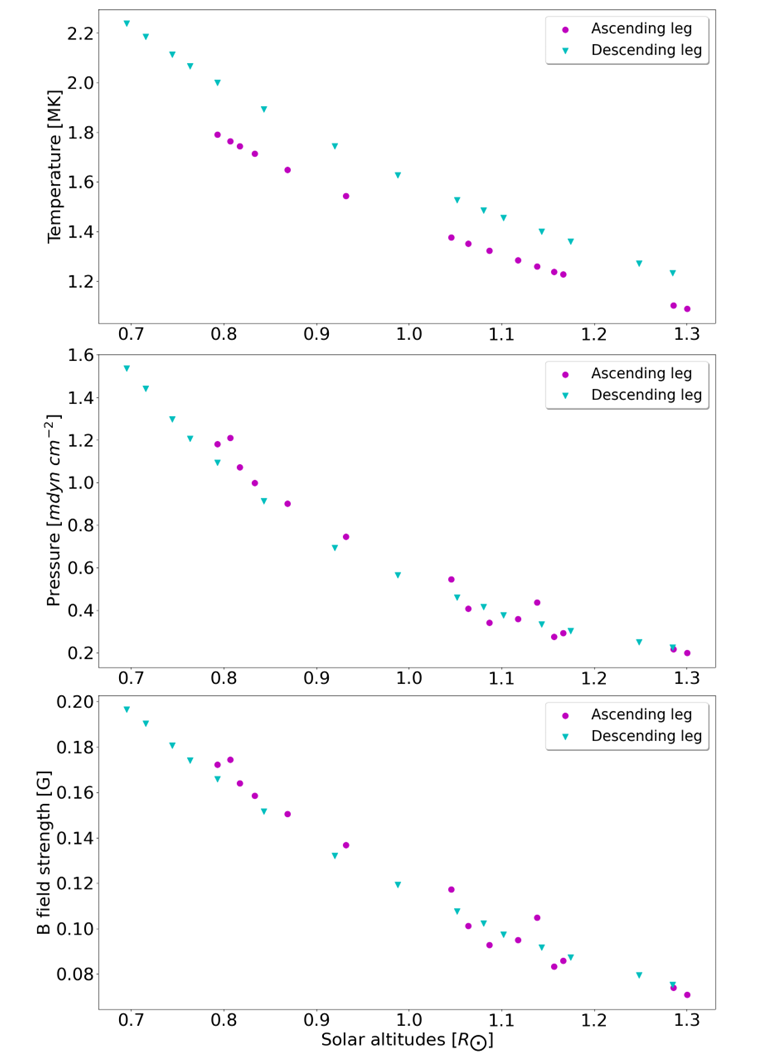}
\caption{
Plasma parameters of the coronal loop derived from the U-burst observations vs the coronal loop altitude. Top: temperature. Middle: pressure.  Bottom: minimum magnetic field strength assuming $\beta<1$). The ascending leg is plotted as purple dots whilst the descending leg is potted as cyan triangles.
\label{fig:ParaD}}
\end{figure}

\subsection{Loop top temperature}\label{sec:Temperature}

Using the coronal loop density scale height, the loop top temperature can be determined by inferring the density model and assuming that the coronal plasma is in hydrostatic equilibrium. We use the hydrostatic density scale height equation \citep[e.g.][]{Aschwanden1992}.  However, due to the large height of our coronal loop, we take into account the change in gravitational acceleration as a function of heliocentric altitude rather than assuming gravitational acceleration is a constant.  Consequently, we do not have an isothermal loop like \citet{Aschwanden1992}.  Instead the loop temperature can be described by:

\begin{equation} \label{eqn:Temperature}
T(r) = \frac{\beta}{1+\alpha} \frac{\mathrm{m_P} \lambda}{\mathrm{k_B}}\frac{\mathrm{GM_\odot}}{r^2},
\end{equation}

\noindent where $\lambda$ is the density scale height estimated from both legs by fitting an exponential model using Equation \ref{eqn:exponential} described in the previous section, and $r$ is the distance from the center of the Sun. The term $\frac{\mathrm{GM_\odot}}{r^2}$ represents the gravitational acceleration, which varies as a function of $r$ due to the large size of the coronal loop. $\alpha=1.22$ is the ratio of electron to proton number, and $\beta=1.44$ is the mean molecular weight, $\mathrm{k_B}$ is the Boltzmann constant, $\mathrm{m_P}$ is the mass of a proton. Therefore, we determined the temperature distributions along both the upward and downward legs of the coronal loop from both the ascending and descending portions of the U-burst as shown in Figure \ref{fig:ParaD}.

On the ascending leg, from the 40~MHz source to the 20~MHz source position, the temperature decreases from 1.76 MK to 1.09 MK. Then, on the descending leg, the temperature increases from the loop top to the lower altitude, ranging from 1.23 MK to 2.24 MK. The temperature near the top of the coronal loop is 1.1 MK on the ascending leg at 1.3 $\rm{R_{\odot}}$, and 1.2 MK on the descending leg at the same altitude. These results are consistent with the estimations by \cite{Zhang2023}, based on a statistical analysis of 24 J-bursts in the same LOFAR LBA frequency range, which suggest an average loop top temperature of 1.0 ± 0.2 MK at an altitude of 1.37 $\rm{R_{\odot}}$.

This supported the plasma temperature is symmetric within the loop. The lower temperature on the ascending leg is due to $\lambda_{descending}= 0.43~\rm{R_{\odot}}$ is greater than $\lambda_{ascending}= 0.38~\rm{R_{\odot}}$ in equation \ref{eqn:Temperature}, as we described in section \ref{sec:DEN&Vel}.

\subsection{Loop top Pressures}\label{sec:Pressure}

The loop top pressure was estimated using the ideal gas law, which incorporates the plasma density ($n_\mathrm{e}$) and loop temperature ($T$) as follows:

\begin{equation} \label{eqn:GASlaw}
P(h) = n_\mathrm{e}(h) \mathrm{k_b} T(h),
\end{equation}

where $T_\mathrm{e}(h)$ is the previously determined loop plasma temperature.

It is important to note that previous studies \citep[e.g.][]{Aschwanden1992} utilized the R-T-V (Rosner-Tucker-Vaiana) law \citep[see][]{Rosner1978} to establish the relationship between pressure, temperature, and loop size. However, this relationship is not applicable in our study due to the condition that the ratio of the density scale height to the loop height is greater than one, which is necessary for using the R-T-V law.

We estimated that the plasma pressure near the loop top at 1.3 $\rm{R_{\odot}}$ is 0.20 $\rm{mdyn~cm^{-2}}$ on the ascending leg and 0.23 $\rm{mdyn~cm^{-2}}$ on the descending leg. These results are lower than the estimation by \cite{Zhang2023} of 0.7 $\pm$ 0.3 $\rm{mdyn~cm^{-2}}$ at 1.37 $\rm{R_{\odot}}$. By using the loop top plasma temperature distribution values determined from section \ref{sec:Temperature}, Figure \ref{fig:ParaD} middle panel illustrates the loop top pressure distributions of both legs. The loop top pressure along the ascending leg, following the path of the electron beam, decreases from 1.2 $\rm{mdyn~cm^{-2}}$ to 0.2 $\rm{mdyn~cm^{-2}}$, and then increases on the descending leg from 0.2 $\rm{mdyn~cm^{-2}}$ to 1.5 $\rm{mdyn~cm^{-2}}$. The 13\% difference at 1.3 $\rm{R_{\odot}}$ is not significant, suggesting both legs have similar plasma pressure. Similar to the temperature, the loop top plasma pressure is lower at higher altitudes, indicating that both legs are similar in this regard.

\subsection{Loop top minimum magnetic field strength}\label{sec:B_field}

In the middle corona, the thermal pressure of the plasma is typically lower than the magnetic pressure, particularly for magnetic loop structures where plasma and particles are 'frozen into' the magnetic fields. This behavior is characterized by the plasma $\beta$. To determine the minimum magnetic field strength of the coronal loop, we utilize the condition where the plasma $\beta$ is less than one:

\begin{equation} \label{eqn:magfield}
B(h) > [8\pi n_\mathrm{e}(h) \mathrm{k_b} T_\mathrm{e}(h)]^{0.5}.
\end{equation}

We determined that the minimum magnetic field strength near the loop top at 1.3 $\rm{R_{\odot}}$ is 0.07 G on the ascending leg and 0.08 G on the descending leg. This is only 20\% lower to the estimation by \cite{Zhang2023} of 0.13 ± 0.03 G at 1.37 $\rm{R_{\odot}}$. Similar to loop top pressure, the distribution of minimum magnetic field strength for both legs is plotted in Figure \ref{fig:ParaD}. The field strength decreases along the ascending leg from 0.17 G to 0.07 G outwards from the 40 MHz source position, and then increases on the descending leg from 0.08 G to 0.20 G downwards from the loop apex.

\section{Discussions}\label{sec:Result_Disc}

\subsection{Apparent Electron Beam Deceleration}

Based on our estimations of electron beam velocities from both legs of the coronal loop, we concluded that the apparent electron beam velocity was less in the descending leg and more in the ascending leg, creating an apparent electron beam deceleration. This decrease in velocity reflects the electron energies that generate Langmuir waves which ultimately result in the detected radio emission we observe. 

Numerous previous simulation studies have demonstrated that the initial properties of the electron beam (e.g. energy distribution, beam density) influence the energy of electrons in the beam which resonate with Langmuir waves \citep{Kontar2001, Li2013a, Li2013b, Li2014, Ratcliffe2014, Reid2018}.  Recently, \citet{Lorfing2023} simulated the generation of Langmuir waves by an electron beam as it propagated from the surface of the Sun to 50 $\rm{R_{\odot}}$. The authors shows that the maximum electron velocity which resonates with Langmuir waves decreases whilst traveling from a high to a low-frequency emission region (low to high altitude). This decreases relates to the energy density of the electrons decreasing as a function of distance, as they spread out along the expanding magnetic flux tube and undergo velocity dispersion \citep{Reid2015,Reid2017}. From observations, a decrease in the apparent velocity has been observed from type III bursts at low frequencies corresponding to the solar wind \citep{Dulk1987,Krupar2015}. 

One key difference in the descending leg of the coronal loop is the sign of the background density gradient: it is positive instead of negative.  A positive density gradient is known to reduce the level of Langmuir waves produced by an electron beam, as refraction shifts waves to higher phase velocities instead of lower phase velocities \citep[e.g.][]{Kontar2001}.  This behaviour is known to reduce the radio intensity of reverse type III bursts \citep{Li2011}.  Whilst the higher phase velocity Langmuir waves could resonate with higher energy electrons, the increased difficulty for the electrons to generate Langmuir waves and the reduced electron flux at higher velocities would likely lead to the apparent deceleration that we have observed here. A further difficulty in the resonant production of Langmuir waves for the descending electron beam is the increasing background plasma temperature and the resultant increase in Landau Damping. Future simulations that tackle this problem and a more robust study of multiple U-bursts that can be imaged are required to investigate this further.

\subsection{Loop Top Plasma Density Model and Physical Parameters}\label{sec:DiscussPara}

Using the solar U-burst, we defined the distribution of electron plasma density of the magnetic loop as a function of solar altitude above the solar surface, as shown in Figure \ref{fig:NRplot}. Both legs of the distribution exhibit similar plasma density distributions, aligning well with Saito's density model. Notably, our results show a lower density model compared to the estimation by \cite{Reid2017}, where Saito's model was multiplied by a factor of 4.5 (also plotted in Figure \ref{fig:NRplot}) when analyzing U-bursts observed by LOFAR between 40 to 80 MHz. This same multiplier of 4.5 was also used by \cite{Zhang2023} in determining beam speeds and deriving loop parameters from J bursts observed between 20 to 80 MHz.  As observed by \cite{Trottet1982}, type III burst can be related to electron beams travelling along discrete, short-lived structures that were denser than the surrounding corona.  Similar results have been found from other studies such as by \citet{Carley2016}, who used 450-150 MHz type III burst imaging to deduce a density model that was 11.5 times that of Saito's density model.  However, combining imaging of 240-80 MHz type III bursts with an MHD model, \citet{McCauley2017} found coronal structures that were similar with the Saito model, similar to our results.  It would be beneficial to use imaging of radio bursts from events at many different times and dates to build up a more statistical picture of how coronal loop density distributions can vary.

The plasma pressure that we found for the coronal loop agrees very well with the estimations of pressure using Yohoh/SXT limb data from \citet{Gary1999}, and shown in \citet{Gary2001} at an altitude of 1.3 solar radii.  The minimum magnetic field strength we estimated using $\beta < 1$ is also within the range of field strengths shown in \citet{Gary2001} at the same altitude.  Our estimate is smaller than the value of the magnetic field estimated by \citet{Dulk1978}, around 0.3 G at the same altitude.  If the plasma beta was 0.1, the minimum value shown in \citet{Gary2001}, we would get a similar value of magnetic field strength to that estimated by \citet{Dulk1978}.

\subsection{Radio Source Size and 
 Scattering Effect}\label{sec:SourceSizeScat}

\begin{figure}[ht!]
\plotone{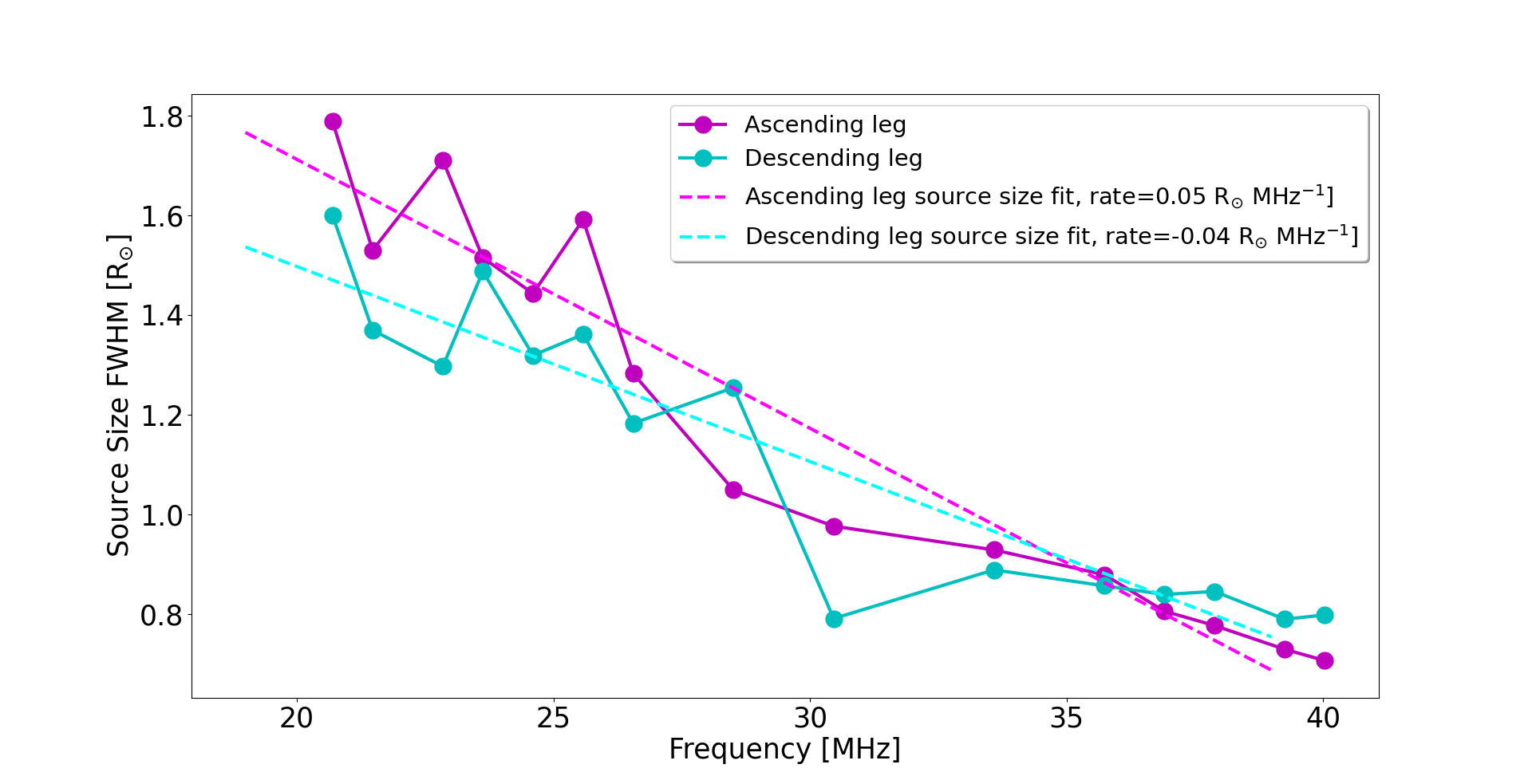}
\caption{
Full Width Half Maximum (FWHM) size of the observed U-burst radio source, defined by the semi-major axis size, for frequencies between 20--40 MHz. The dashed lines are linear fits to the data. 
\label{fig:SourceSize}}
\end{figure}

The size of the radio source is an important measurement that can be obtained from solar radio imaging. By examining the interferometric image of the U-burst in Figure \ref{fig:TheImg}, we plotted the radio source size, defined by the semi-major axis length of the Full Width Half Maximum (FWHM), as a function of frequency shown in Figure \ref{fig:SourceSize}. The size of the radio source increased while the electron beam was traveling along the ascending leg and decreased while it was traveling along the descending leg. We discovered that the radio source expanded and contracted at a similar rate on both legs: the ascending leg source size had an increase rate of 0.5 $\rm{R_{\odot}}~\rm{MHz^{-1}}$, and the descending leg had a decrease rate from the loop top, propagating downward at a rate of -0.4 $\rm{R_{\odot}}~\rm{MHz^{-1}}$. 

From the U-burst image displayed in Figure \ref{fig:TheImg}, it is evident that the coronal loop's cross-section is larger at the loop top and gradually decreases towards the footpoint of the leg as shown in Figure \ref{fig:SourceSize}. This loop geometry aligns with the classification of coronal loops by \cite{Zaitsev2017}, which indicates that the plasma pressure within the loop is high, and the flux tube cross-section increases with height. The expansion of the cross section of the coronal magnetic flux tubes leads to a decrease in the electron beam density, which subsequently reduces the generation of Langmuir waves. As a result, the radio emissions in the lower frequency part of type-III bursts and the curvature part of U bursts commonly become more diffused. This phenomenon has been discussed and studied in the work by \cite{Reid2015,Reid2017}, related to type III and U-burst stopping frequencies. 

Other than the magnetic tube structural expansion, another factor that significantly influences the observed size and shape of the radio source is the radio wave propagational scattering effect, as demonstrated in numbers of previous studies \citep[for example,][]{Steinberg1971, Bian2019, Sharykin2018, Kontar2017, Kontar2019,  Kontar2023}. This scattering effect, induced by anisotropic density turbulence, causes the size of the intrinsic source to be observed as a larger radio source by Earth-based radio telescopes.  The 35 MHz radio source of both legs has a size around 0.87 $\rm{R_{\odot}}$, close to \cite{Kontar2019}'s calculation of 1.1 $\rm{R_{\odot}}$ based upon the scattered size of a point source, assuming fundamental emission.  Moreover, we also find that a $f^{-1}$ dependence to the source size with frequency is a reasonable fit to the data, a dependence that is predicted from radio wave scattering.  As such, we can conclude that scattering is likely to play a significant role in the determination of the U-burst radio source sizes.

Another effect of radio wave scattering is the outward shift of radio sources on the image away from the solar center displaced from their actual source location. This means the actual source is at lower altitude than the location on the U-burst image. In this case, the loop temperature estimations would be higher as the gravitational acceleration in Equation \ref{eqn:Temperature} becomes larger, so the pressure and the minimum magnetic field strength would be greater as well.  We identified that the emission mechanism of the U-burst flux is second-harmonic emission, which involves less intrinsic shift of the source position compared to emission at the fundamental, and so our results will be less affected.  With respect to our estimated velocities, this derived parameter is related to the relative distance between each radio source contour and not the absolute position.  Consequently, scattering effects will have a very limited influence on the estimations of the electron beam velocities.

\subsection{Projection Effect}\label{sec:Projection}

The projection effect in solar imaging presents a significant challenge when measuring the distances of radio sources along the line-of-sight axis (z-axis). This issue arises because 2D projected radio images only yield positional information on the x and y axes. The projection effect substantially influences the estimation of electron beam velocities and the assumed density model. It also introduces uncertainties in accurately determining both the actual electron beam acceleration site and the descending leg’s footpoint of the loop.

In this study, we made the assumption that the ascending leg's footpoint of the coronal loop is located at Active Region AR12765, based on simultaneous observations of a coronal jet. We posited that the ascending leg aligns with the same longitudinal plane as AR12765 (at -60° longitude), and we presumed the descending leg to be on the plane of the solar limb (at -90° longitude). Distances along the z-axis for each radio source were measured by incorporating the observational angle derived from this assumption. Therefore, the descending leg radio source heliocentric altitude determined based on this assumption is the underestimation of the real situation, as the z-axis distances are zero on the limb plane. \citet{Chen2023} analyzed the interplanetary component of the type III bursts that we imaged during our study.  The kHz radio source positions, deduced using  multiple spacecraft observations were found to have a good agreement with the 60 degree Parker spiral that we found for the origin of the type III bursts and U-bursts, detected at higher frequencies with LOFAR.

\section{Conclusions}\label{sec:Conclusions}

In this study, for the first time, we used a radio U-burst to successfully image both the upward and downward legs of a large coronal loop extending into the middle corona.  The U-burst was part of a series of radio bursts that lasted 90 seconds.  The exciting electron beams were accelerated in an active region on the southern hemisphere, accompanied by a EUV jet.  By employing high temporal and frequency resolution interferometric imaging data provided by LOFAR, we determined the loop altitude was around 1.3 $\rm{R_{\odot}}$ above the solar surface, originating in an active region in the southern hemisphere and extending into the northern hemisphere.

We inferred density models of the coronal loops for both legs and derived the physical parameters of the coronal loop based on the density scale height from the inferred model, which is 0.38 $\rm{R_{\odot}}$ for the ascending leg and 0.43 $\rm{R_{\odot}}$ for the descending leg. The loop top part plasma temperatures, pressures, and minimum magnetic field strengths were found to be similar on both legs. At 1.3 $\rm{R_{\odot}}$, where the 21 MHz sources on both legs are located, the plasma temperature to be around 1.1 MK, the plasma pressure around 0.20 $\rm{mdyn,cm^{-2}}$, and the minimum magnetic field strength around 0.07 G. The physical parameters distribution along the loop top part determined from both legs are similar, as Figure \ref{fig:ParaD} illustrates. Therefore, we consider this as evidence that the large coronal loop possesses symmetric physical properties. The density model that we inferred from the U-burst was very similar to the Saito density model.  Assuming this density model in \citet{Zhang2023} would result in lower altitudes and consequently higher temperatures, pressures and magnetic field strengths for the derived coronal loops. 
 Although U-bursts with such clear and wide frequency range are unusually seen, we suggest that more observations and comparisons are needed in the future to further explore the conclusion of symmetry of physical parameters of both upflow and downflow legs of such large-sized coronal loops.

We determined the velocity of the electron beam that excited the radio U-burst, based on the ascending and descending leg radio source positions. Our findings reveal an apparent deceleration in the electron beam propagation along the coronal loop, with velocities decreasing from 0.21c to 0.14c. As discussed in Section \ref{sec:Velocity}, we propose a number of physical parameters may many a role in this apparent deceleration, including the magnetic field expansion, velocity dispersion, and the change in the sign of the background density gradient that occurs at the top of the magnetic loop, making it more difficult for high velocity electrons within the beam to generate Langmuir waves that can produce the observable radio waves \citep[e.g.][]{Kontar2001,Reid2015}.  We note that we do not think there was any deceleration of the electron beam, merely that a decreasing energy range of electrons resonated with the Langmuir waves as a function of distance \cite[e.g.][]{Lorfing2023}.

We examined the impact of both scattering and projection effects on radio imaging. We acknowledge that the actual radio source positions could be lower than the observed altitudes due to the scattering effect, as discussed by \citet{Kontar2023}. However, considering the emission mechanism of the U-burst is second-harmonic, we propose that the scattering effect's influence might be less significant compared to that of fundamental emissions. Furthermore, the scattering effect is unlikely to significantly affect the estimation of electron beam velocities and the density scale heights in this study, as the relative distances between each imaged radio source are less impacted than the absolute heights. The projection effect remains a challenge for ground-based solar radio imaging between 20 to 80 MHz. Therefore, we advocate for future observation in this range by multiple instruments from different observational positions to mitigate this issue.

\begin{acknowledgments}
 
This article is based on data obtained with the International LOFAR Telescope (ILT), e-Callisto Solar Spectrogram data, the NDA/NewRoutine and Mefisto Sun dataset, and SDO AIA, Sunpy, PFSS. LOFAR \citep{vanHaarlem2013} is the Low Frequency Array designed and
constructed by the Netherlands Institute for Radio Astronomy (ASTRON). The authors thank the staff of ASTRON and the LOFAR KSP group. JZ acknowledges the use of The Default Pre-Processing Pipeline (DPPP) and WSclean \citep{offringa-wsclean-2014} for producing interferometric solar radio images in this research.  The authors acknowledge Fachhochschule Nordwestschweiz (FHNW) Campus Brugg/Windisch and Istituto Ricerche Solari (IRSOL) Locarno, Switzerland, for the use of e-Callisto Solar Spectrogram data in this study. 

JZ acknowledges the Royal Astronomical Society (RAS) and the Dublin Institute for Advanced Studies (DIAS) for providing travel funds for necessary travels and computational resources for generating LOFAR solar radio images. JZ acknowledges the Mullard Space Science Laboratory (MSSL) and ASTRON for supporting analyses of LOFAR interferometric data.

HASR acknowledges funding from the STFC Consolidated Grant ST/W001004/1. Both JZ and HASR acknowledge support from the Royal Society International Exchange Project IEC$\backslash$R2$\backslash$202175.

The authors acknowledge the Nançay Radio Observatory / Unité Scientifique de Nan\c cay of the Observatoire de Paris (USR 704-CNRS, supported by Universit\'e d’Orl\'eans, OSUC, and R\' egion Centre in France) for providing access to NDA observations accessible online at \url{https://www.obs-nancay.fr}. The NDA/NewRoutine and Mefisto Sun dataset are distributed under the CC-BY 4.0 licence and referenced as \citep{NDA-Sun-NewRoutine,NDA-Sun-Mefisto}. The french co-authors acknowledge support from CNRS/INSU national programs of Planetology (PNP) and Heliophysics (PNST). 

\end{acknowledgments}

\bibliography{bib}{}
\bibliographystyle{aasjournal}



\end{document}